\documentclass[12pt]{article}
\usepackage[pctex32]{graphics}
\begin{document}

\vspace{2cm}

\begin{center} {\large \bf Fuzzy Rings in  D6-Branes and Magnetic Field Background}
                                                  
\vspace{1cm}

                      Wung-Hong Huang\\
                       Department of Physics\\
                       National Cheng Kung University\\
                       Tainan,70101,Taiwan\\

\end{center}
\vspace{1cm}
\begin{center} {\large \bf  Abstract} \end{center}
We use the Myers T-dual nonabelin Born-Infeld action to find some new nontrivial solutions for the branes in the background of  D6-branes and Melvin magnetic tube field.   In the  D6-Branes background we can find both of the  fuzzy sphere and fuzzy ring solutions, which are formed  by the gravitational dielectric effect.   We see that the fuzzy ring solution has less energy then that of the fuzzy sphere.   Therefore the fuzzy sphere will decay to the fuzzy ring configuration.   In the Melvin magnetic tube field background there does not exist fuzzy sphere while the fuzzy ring configuration may be formed by the magnetic  dielectric effect.   The new solution shows that $D_0$  propagating in the D6-branes and magnetic tube field background may expand into a rotating fuzzy ring.  We also use the Dirac-Born-Infeld action to construct the ring configuration from the D-branes.

\vspace{1cm}
\begin{flushleft}
E-mail:  whhwung@mail.ncku.edu.tw\\
\end{flushleft}
\newpage
\section  {Introduction}

The realization that the D-branes play a fundamental role in the understanding of string and M-theory has played  a central role in the developments of the past years [1,2].   It is known that higher-dimensional branes may be constructed from the lower-dimensional branes.  Especially, some nontrivial configurations, such as the fuzzy sphere, supertubes, torus, cylinder, and plane  may be formed under the dielectric effect [3,4].   

In recent, the investigations have been extended to the theory in the curved spacetime.  Among these the spacetime with Melvin metric [5,6] has been studied extensively.   Melvin metric is a solution of  Einstein-Maxwell theory, which describes a static spacetime with a cylindrically symmetric magnetic flux tube.   It provides us with a curved space-time background in which the superstring theory can be solved exactly [7].   In the Kaluza-Klein spacetime the Melvin solution is a useful metric to investigate the decay of magnetic field [8] and the decay of spacetimes, which is related to the closed string tachyon condensation [9].   The fluxbranes in the Melvin spacetime have many interesting physical properties as investigated in the resent literatures [10].

   In [11] Hyakutake had investigated the gravitational dielectric phenomena of a D2-brane in the background D6-branes.  He has shown that the spherical D2-brane with nonzero radius becomes classical solution
of the D2-brane action.   The corresponding curved background spacetime is the classical solution of $N$ coincident D6-branes, which was found by Gibbons and Maeda [4].  

   In this paper we try to investigate the branes dynamics in the background of  D6-branes and Melvin magnetic tube field by the nonabelin Born-Infeld action.  The Myers T-dual nonabelin Born-Infeld action adopted in this paper is given by [3].   
$$S_{BI}=-T_p \int d^{p+1}\sigma\, Tr\left(e^{-\phi}\sqrt{-det\left(P\left[E_{ab}+E_{ai}(Q^{-1}-\delta)^{ij}E_{jb}\right]+ \lambda F_{ab}\right)\, det(Q^i{}_j)} \right) ,   \eqno{(1.1)}$$
in which 
$$ E_{kj} =G_{kj}+B_{kj},    \eqno{(1.2)}$$
$$ Q^i{}_j\equiv\delta^i{}_j+{i\over \lambda}~ [X^i, X^k]~ E_{kj},  \eqno{(1.3)}$$
and pull-back of the bulk spacetime tensors $G_{MN}$ to the D-brane world-volume is denoted by the symbol $P[G_{MN}]$,  
$$ P[G]_{ab}=G_{ab}+ ~G_{i(a}\partial_{b)} X^i+ G_{ij}\partial_a X^i\partial_b X^j . \eqno{(1.4)}$$
The brane tension is defined by
$$ T_p={2\pi\over g_s\left(2\pi\ell_s\right)^{p+1}}, \eqno{(1.5)}$$
and  $\lambda \equiv 2\pi \ell_s^2$ in which $\ell_s$ is the string length scale.  In section II we use the above formula to evaluated the nonabelin Born-Infeld action in the background of D6-branes.  We show that besides the fuzzy sphere solution, which was found by Hyakutake [11],  there is also the fuzzy ring solution (The definition of fuzzy ring is made in comment 1 of section 2.).   We also see that the ring solution has lower energy then  that of the sphere solution.  In section III we formulate the nonabelin Born-Infeld action in the background of Melvin magnetic tube field.  We show that there does not exist the fuzzy sphere solution.   However, in this case we still have the fuzzy ring solution.   We also extend the investigation to the nonstatic system.    The results show that the $D_0$  propagating in the D6-branes and magnetic tube field background may expand into a rotating fuzzy ring.  (Note that in a previous letter we had also found the rotating fuzzy ring  in the Melvin Matrix model [13,14].)  In section IV we also use the Dirac-Born-Infeld action [12] to investigate the problems of  constructing  the ring from the D-branes.   Last section is devoted to a discussion.

\section {Fuzzy Ring in D6-branes Background}

The spacetimes of the classical solution of $N$ coincident D6-branes is given by the metric 
 $$ds^2 = f^{-\frac{1}{2}} \eta_{\mu\nu}dx^\mu dx^\nu 
  + f^{\frac{1}{2}} \left(\delta _{ij} dx^i dx^j\right),  \eqno{(2.1)}$$
$$e^\phi = f^{-\frac{3}{4}} ,  ~~~  f = 1 + \frac{N \ell_s g_s}{2r} ,  ~~~ G^{(2)} = \frac{N\ell_s g_s}{2} \sin\theta d\theta \wedge d\phi, \eqno{(2.2)}$$
where $\mu,\nu = 4,\cdots,9,0$, which label the tangent directions to D6-branes and $i,j=1,2,3$, which describe normal ones. The radius $r$ is given by $r = \sqrt{\delta_{ij}x^i x^j}$.   $G^{(2)}$ is the R-R 2-form flux originating from the D6-branes. This flux plays no role in the discussions below.  (See the comment 3 in this section)

  To proceed, we first set $X^4 = \cdots = X^9 = 0$ as we are interested in the configurations of D0-branes which form a fuzzy geometry in the transverse space of D6-branes.   Then the potential energy of  M static D0-branes in the background of  $N$ coincident D6-branes (2.1)  can be evaluated by (1.1).   The result is 
$$ V_{D0} = T_0 ~ STr \bigg( \sqrt{ f - \frac{f^2}{2\lambda^2}
  [X^i,X^j]^2 } \, \bigg) . \eqno{(2.3)}$$
In this case, for the following physical interpretation, we can express the radius $r$ in terms of $X^i$ as 
$$r = \sqrt{\frac{1}{M} {Tr} (X^i)^2}.  \eqno{(2.4)}$$ 
In case of $f \ll - \frac{f^2}{2\lambda^2}[X^i,X^j]^2$, the potential energy of M D0-branes expanded to the quadratic order becomes 
$$ V_{D0} \sim T_0 ~ f^{1/2} STr \left( 1 - \frac{f}{4\lambda^2}
  [X^i,X^j]^2  \right) $$
 $$ \sim a \, r^{-\frac{1}{2}} - b \, r^{-\frac{3}{2}}
  Tr~ \big( [X^i,X^j]^2 \big) , \eqno{(2.5)}$$
where 
$$a \equiv MT_0 ( \frac{N\ell_s g_s}{2} )^{\frac{1}{2}}, \eqno{(2.6a)}$$
$$b \equiv \frac{T_0}{4\lambda^2} ( \frac{N\ell_s g_s}{2} )^{\frac{3}{2}},\eqno{(2.6b)}$$
and STr represents the symmetrized trace prescription [3].  Note that to obtain the last relation in (2.5) we have used the approximation $f = 1 + \frac{N \ell_s g_s}{2r} \approx \frac{N \ell_s g_s}{2r}$ as we consider only the region near the D6-branes.  (It is worth noting that in the D6-brane case,  the 6+1 dimensional  gauge theory on the D6-branes, unlike the lower dimensional branes, is not decoupled from the bulk gravity [15].)

    Now  we can use the relation $\frac{\delta r}{\delta X^i} = \frac{1}{Mr} X^i$ to obtain the equations of motion  
$$   \frac{\delta V_{D0}}{\delta X^i}  \sim  \frac{r^{-\frac{5}{2}}}{2M} \Big\{ - aX^i + 3 b r^{-1}X^i   Tr~ \big( [X^j,X^k]^2 \big) - 8Mb r [X^j,[X^i,X^j]] \Big\} = 0.  \eqno{(2.7)}$$
There are two types solution in the above equation.

   1. {\it \bf Fuzzy sphere solution}:   The first solution is given by
$$    [X^i,X^j]= \pm i\epsilon^{ijk}\frac{2r_\ast}{\sqrt{M^2-1}}X^k, \eqno{(2.8)}$$
$$r_\ast = \left(\frac{aM(1-M^{-2})}{40b}\right)^{\frac{1}{3}}.  \eqno {(2.9)}$$
The solution represents the fuzzy sphere with the radius $r_\ast$. The Myers effect in here is purely gravitational.   This is that first found in [11] by Hyakutake.  Substituting the fuzzy sphere into (2.5) we see that the configuration has the energy 
$$V_{sphere}=  a \left[{a (M^2-1)\over 40 b M}\right]^{-1/6} +  {8bM\over M^2-1} \left[{a (M^2-1)\over 40 b M}\right]^{5/6}. \eqno{(2.10)}$$

   2.  {\it \bf Fuzzy ring solution}:   Another new solution we found is described by  
$$ X^1= \left({a \sqrt 3\over b M \sqrt {M^2 -1}}\right)^{1/3} J_x,~~~~~ X^2= \left({a \sqrt 3\over b M \sqrt {M^2 -1}}\right)^{1/3} ~ J_y, ~~~~~~ X^3=0, \eqno{(2.11)}$$
in which $J_i$ are the $M \times M$ representation of $SU(2)$.  The solution represents the fuzzy ring with the radius 
$$r_\ast = \left(\frac{aM(1-M^{-2})}{30b}\right)^{\frac{1}{3}},  \eqno {(2.12)}$$
following the definition in (2.4).   Now, substituting the solution into (2.5) we see that the fuzzy ring configuration has the energy 
$$V_{ring}=  a \left[{a (M^2-1)\over 30 b M}\right]^{-1/6} +  {6bM\over M^2-1} \left[{a (M^2-1)\over 30 b M}\right]^{5/6}. \eqno{(2.13)}$$
Comparing (2.10) with (2.13) we find that 
$${V_{sphere}\over V_{ring}} = \left({4\over3}\right)^{1/6} > 1, \eqno{(2.14)}$$
and the fuzzy ring solution has less energy then that of the fuzzy sphere.   Therefore the fuzzy sphere is metastable and would decay to the fuzzy ring configuration.

  We make some comments to conclude this section.

1.  A fuzzy sphere is a $S_2$ configuration in which the coordinates $X^1$, $X^2$, and $X^3$ are proportional to $J_x$, $J_y$, and $J_z$ respectively ($J_i$ are the representation of $SU(2)$).  While $Tr \left(J_x^2 + J_y^2 + J_z^2\right)$ is a constant which is proportional to the spherical radius (see (2.9)) the spherical coordinates $X^1$, $X^2$, and $X^3$ do not commute each other.  It is this property that we call the configuration a fuzzy sphere.  In the same way, a fuzzy ring is a $S_1$ configuration in which the coordinates $X^1$ and $X^2$ are proportional to $J_x$ and $J_y$ respectively.  (Note that $X^3 = 0$.)  While $Tr \left(J_x^2 + J_y^2\right)$ is a constant which is proportional to the ring radius (see (2.12)) the ring coordinates $X^1$ and $X^2$ do not commute each other.   It is this property that we call the configuration a fuzzy ring.

2. It shall be mentioned that not every system which has fuzzy sphere could also have  fuzzy ring solution.  For example, the original matrix model with Chen-Simon term [3] have the well known fuzzy sphere solution, but, as can be easily seen, the model does not contain the fuzzy ring solution.   However, the systems in the background of  D6-branes have both of  fuzzy sphere and fuzzy ring configurations, as shown in this section.

3.  To prove that the RR 2-form flux  $G^{(2)}$ originating from the D6-branes does not affect the ring solution let us investigate the value of $det(Q^i{}_j)$ in (1.1).  RR 2-form flux defined in (2.2) implies that 
$$B_{xy} = {{N \ell_s g_s}\over \sqrt{x^2+y^2+z^2}} {-z/2\over x^2+y^2+Z^2}, \eqno{(2.15a)}$$
$$B_{yz} = {{N \ell_s g_s}\over \sqrt{x^2+y^2+z^2}} {-x/2\over x^2+y^2}, \hspace{1cm}\eqno{(2.15b)}$$
$$B_{zx} = {{N \ell_s g_s}\over \sqrt{x^2+y^2+z^2}} {-y/2\over x^2+y^2}. \hspace{1cm}\eqno{(2.15c)}$$
Using the above values of $B_{MN}$ and the metric $G_{MN}$ in (2.1) the values of $E_{MN}$ in (1.2) can be found.   Then, through the calculation we have a simple relation
 $$det(Q_i^j) = 1 - \frac{f}{4\lambda^2}[X^i,X^j]^2 - \frac{1}{\lambda^2}\left( B_{xy}[X^1,X^2] + B_{yz}[X^2,X^3]+B_{zx}[X^3,X^1]  \right)^2. \eqno{(2.16)}$$
The ring solution of (2.11)  has the properties $X^3 = 0$ and $z=0$. Thus from (2.15a) we see that $B_{xy} = 0$.   Therefore the third term in (2.16) is vanishing and  RR 2-form flux $B_{MN}$ in the form (2.2) does not affect the ring solution.   Because of this property and considering only the static system, i.e., $\dot X^i =0$ the action (1.1) does not be changed if the RR 2-form flux with the form  (2.2) is introduced. (Note that in (2.16) we have neglected the higher order terms of $[X^i,X^j]^n$, $n>2$.)

4. It is interesting to see that the action has clearly an SU(2) symmetry, rotating $X^i$  into each other, and generally one would physically expect the vacuum state (the minimum of the potential) to have the same symmetry which is, however, apparently absent in the ring solution.  In fact, besides the solution of (2.11) we also have following two solutions  
$$ X^2= \left({a \sqrt 3\over b M \sqrt {M^2 -1}}\right)^{1/3} J_y,~~~~~ X^3= \left({a \sqrt 3\over b M \sqrt {M^2 -1}}\right)^{1/3} ~ J_z, ~~~~~~ X^1=0, \eqno{(2.17)}$$
$$ X^1= \left({a \sqrt 3\over b M \sqrt {M^2 -1}}\right)^{1/3} J_x,~~~~~ X^3= \left({a \sqrt 3\over b M \sqrt {M^2 -1}}\right)^{1/3} ~ J_3, ~~~~~~ X^2=0. \eqno{(2.18)}$$
The three solutions have the same energy and choosing one of the solution as the vacuum state will spontaneously break the system symmetry from SU(2) to U(1).

    In the next we will investigate nonabelin Born-Infeld action  in the background of  Melvin magnetic tube field.  We will see that in the Melvin magnetic tube field background there does not exist fuzzy sphere and fuzzy ring solution can be formed by the magnetic  dielectric effect.   We will also extend the investigation to the nonstatic case and see that $D_0$  propagating in the magnetic tube field background may expand into a rotating fuzzy ring.  

\section  {Fuzzy Ring in Melvin Magnetic Tube Background}

The Melvin spacetime is a solution of the Einstein-Maxwell theory [5].  The Kaluza-Klein Melvin spacetime may be described as a flat manifold subject to non-trivial identifications [8].   Explicitly,  the 11-dimensional flat metric in M-theory is written in cylindrically coordinates
$$ds^2_{11}=-dt^2+ \sum_{m=1}^{7}dy_m dy^m +dr^2+r^2  d\varphi ^2 + dx_{11}^2 ,  \eqno{(3.1)} $$
with the identifications
   $$(t,y_m ,r,\varphi, x_{11}) \equiv (t,y_m,r,\varphi+2\pi n_1 R B +2\pi n_2, x_{11}+2\pi  n_1R), \eqno{(3.2)} $$
The identification under shifts of  $2\pi n_2$ for $\varphi$ and $2\pi  n_1R$ for $x_{11}$ are, of course, standard. The new feature is that under a shift of $x_{11}$, one also  shifts $\varphi$ by $2\pi n_1 R B$.   We can introduce the new coordinate $\tilde\varphi=\varphi-Bx_{11}$ and rewrite the metric as 
$$ds^2=-dt^2+ \sum_{m=1}^{7}dy_m dy^m +dr^2+r^2  (d\tilde\varphi+Bdx_{11}) ^2 + dx_{11}^2 ,  \eqno{(3.3)} $$
with the points $(t,y_m ,r,\tilde\varphi, x_{11})$ and $(t,y_m,r,\tilde\varphi+2\pi n_2, x_{11}+2\pi  n_1R)$ identified. 

    Since the eleven-dimensional spacetime is flat this metric is expected to be an exact solution of the M-theory including higher derivative terms. We can recast the eleven-dimensional metric in the following canonical form  [8]
$$ ds_{11}^2= e^{-2\phi/3}ds_{10}^2+  e^{4\phi/3} (dx_{11}+2 A_\mu dx^\mu )^2 , \eqno{(3.4)}  $$
the ten-dimensional IIA background is then described by
$$ ds_{10}^{2} = \Lambda^{1/2}\left(-dt^2+\sum_{m=1}^{7}dy_mdy^m+dr^2\right)   +\Lambda^{-1/2}r^2d\tilde{\varphi}^2 ,\eqno{(3.5)} $$
$$  e^{4\phi/3}=\Lambda \equiv 1+r^2B^2 ,~~~~ A_{\tilde{\varphi}}=\frac{Br^2}{2\Lambda}.  \eqno{(3.6)} $$
The parameter $B$ is the magnetic field along the $z$-axis defined by
$B^2=\frac{1}{2}F_{\mu\nu}F^{\mu\nu}|_{r=0}$.  The Melvin spacetime is therefore an exact solution of M-theory and can be used to describe the string propagating in the magnetic tube field background [7].

  To proceed we first express the metric (3.5) in the rectangular coordinates
$$ ds_{10}^{2} = \Lambda^{1\over2}\Bigg(-dt^2+\sum_{m=1}^{7}dy_mdy^m\Bigg)  +\Lambda^{-1\over ~2}\Bigg((1+B^2x^2)dx^2 +(1+B^2y^2)dy^2 + 2B^2xy~dxdy\Bigg).\eqno{(3.7)} $$
The magnetic field tensor becomes
$$F_{xy} = {-B\over (1+B^2r^2)^2},\eqno{(3.8)} $$
in which $r^2=x^2 + y^2$.    Now we can set $X^1 = \cdots = X^7 = 0$ as we are interested in the configurations of D0-branes which form a ring configuration.   Then the potential energy of  M static D0-branes in the background of  Melvin magnetic tube field (3.5) can be evaluated by (1.1).   In case of $1 \ll - \frac{1}{2\lambda^2}[X, Y]^2$, the potential energy of  M D0-branes expanded to the quadratic order becomes   
$$V_{D0} = T_0 ~\Lambda^{-1/4} \left(1+ \Lambda^{-4}B^2 \right) \left(M- {1\over 2\lambda^2} Tr [X,Y]^2 \right). \eqno{(3.9)}$$
We can try to find the fuzzy ring by the following ansatz
$$ X= A J_x,~~~~~ Y= A~ J_y.  \eqno{(3.10)}$$
In this case as the radius is defined by 
$$r = \sqrt{\frac{1}{M} Tr(X^2 + Y^2)},  \eqno{(3.11a)}$$ 
and we have the relation 
$$A = \sqrt{6 r^2\over M^2 -1}.  \eqno{(3.11b)}$$ 
Therefore the potential (3.9) becomes
$$V_{D0} = T_0 ~\Lambda^{-1/4} \left(1+ \Lambda^{-4}\lambda^2 B^2 \right) \left(1+{1\over 2\lambda^2} {3r^4\over M^2-1}\right). \eqno{(3.12)}$$
After plotting the potential (3.13) with respect the variable $r$ we can see that the potential has a minimum at a  fine value of $r_\ast$, which is the radius of the fuzzy ring, as shown in figure 1.      For the case of small magnetic field we have a simple relation
$$ r_{\ast} \approx {\sqrt 3\over6}|B|\lambda (M^2-1),~~~~~~~if ~~|B|<<1. \eqno{(3.13)}$$
Thus the radius of the ring is proportional to strength the magnetic field. Note that in the system with of  $[X,Y] = c$, in which $c$ is any constant, the potential becomes minimum at $r\rightarrow \infty$.   This means that the commutative system will runaway to infinite and M D0 may condense into a fuzzy ring configuration with finite radius.
\vspace{1cm}

\hfil\scalebox{1}{\includegraphics{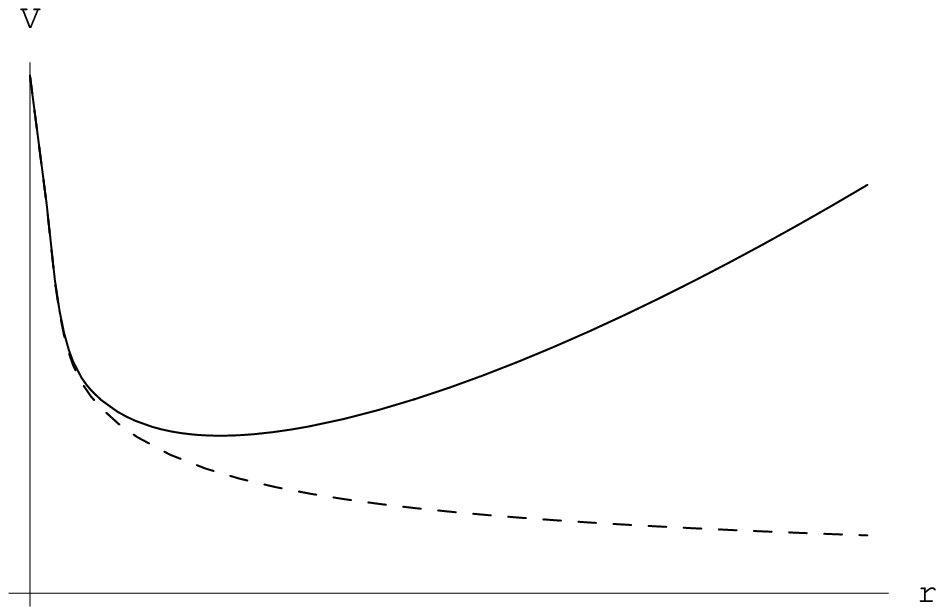}}\hfil\\
{\it ~~~Fig.1. Radius dependence of  the potential (3.12).  The solid line represents the potential of ring which has a minimum.  The dot line is the potential of the trivial solution which does not have minimum at finite radius.}
\vspace{1cm}

    We now turn to investigate the nonstatic system.    In this case the pull-back of the bulk spacetime tensors is 
$$P[E_{ab}]_{tt} = G_{00} + G_{IJ} \dot X^I \dot X^J ,\eqno{(3.14)}$$  
$$P\left[E_{ab}+E_{ai}(Q^{-1}-\delta)^{ij}E_{jb}\right]_{tt} =  \dot X^I ~ G_{Ii}~(Q^{-1})^{ij}~G_{jJ}~ \dot X^J -  G_{IJ} \dot X^I \dot X^J,\eqno{(3.15)}$$
\\
in which the metric $G_{IJ}$, with $I,J = 8,9$ is defined in (3.7) and 

$$ Q^I_{{}J} =\left[ \begin{array}{cc} 
1+ {i\over \lambda} \Lambda^{-1/2}B^2 xy [X, Y]~~~  & {i\over \lambda} \Lambda^{-1/2}(1+B^2 y^2) [X, Y] \\
-{i\over \lambda} \Lambda^{-1/2}(1+B^2 x^2) [X, Y]~~~ & 1- {i\over \lambda} \Lambda^{-1/2}B^2 xy [X, Y] \
\end {array}\right].  \eqno{(3.16)}$$
\\
As the formula in the nonstatic system becomes too complex we will consider the system in the small magnetic field.  The analytic Hamiltonian becomes 
$$ H=  T_0 ~\Lambda^{-1/4} \left(1+ \Lambda^{-4}\lambda ^2B^2 \right) Tr \left[\left(1- {1\over 2\lambda^2} Tr [X,Y]^2 \right)\left(1+  {\Lambda^{1/2}\over 2} (\dot X^2 + \dot Y^2) \right)\right]. \eqno{(3.17)}$$
To find an nonstatic solution from the Hamiltonian (3.17) we can make  the following ansatz
$$X= x(t)~{J_x} =  R cos(f(t)) ~{J_x},\eqno{(3.18)}$$
$$Y = y(t)~{J_y} = R sin(f(t)) ~ {J_y},\eqno{(3.19)}$$
in which $R$ is a constant value and  $f(t)$  a time-dependent function to be determined.  In this ansatz the Hamiltonian (3.17) becomes
$$H_{D0} = T_0 \left(1+{B^2 R^2\over 4}\right)(1+ \lambda^2 B^2) \left[1+{3 R^4\over \ 4 \lambda^2 (M^2-1)}~ sin(2f(t))^2~ + {1\over 4}\left(2+B^2 R^2\right) R^2 \dot f(t)^2 \right]$$
$$ = \alpha (R^2, B^2) \left(\dot {f(t)}\right)^2 + \beta (R^2,B^2) \left(sin(2f(t))\right)^2 + \gamma (R^2,B^2), \hspace{3cm}\eqno(3.20)$$
in which the detailed form of  the functions $ \alpha, \beta$ and  $\gamma $ are irrelevant to the following analyses.  

    The equation (3.20) may be regarded as a particle with mass $2 \alpha (R^2, B^2)$ moving under the potential $\beta (R^2,B^2) \left(sin(2f(t))\right)^2$.   As the Hamiltonian does not explicitely  dependent on time this is an energy-conservation system.  Thus, depending the initial energy, the particle may be oscillating or moving to infinite.  (Note that the coordinate of  the particle motion is the function $f(t)$.)  As the radius $R^2$ ($\equiv x(t)^2+y(t)^2$) is a constant the solution therefore represents a ring which does not change the shape.   The ring is called a fuzzy ring as the coordinates $X$ and $Y$ in (3.18) and (3.19)  are the matrices which do not commutate to each other as $[X, Y] \sim J_z$.  Also, as $x(t)$ and $y(t)$ is time dependent, the fuzzy ring will be rotating.   Depending on the function of $f(t)$, (which is  dependent of  the $B$ and $R$)  the ring may be rotating in righthand or in lefthand.   It may also be rotating form righthand to lefthand or form lefthand to righthand.   

    Along the same investigations it can be seen that the rotating fuzzy ring could also be shown in the D6-Brane background. Note that without the magnetic field the rotating membranes had been found by Harmark and Savvidy [16].   

     Finally we shall mention that in a previous letter [13] we have uses the regularized theory of light-cone gauge memberbranes [14] to find the matrix model in the Kaluza-Klein Melvin background.   We had also in [13] found the rotating fuzzy ring  in the Melvin Matrix model.  However, the metric adopted there is the eleven-dimensional metric in (3.4).   The investigation in this paper is using the nonabelin Born-Infeld action in the ten-dimensional IIA background which has the metric described by (3.5). 

\section  {Construction of  Ring Configuration}
In this section we will investigate how to construct the ring configuration from the D-branes.  As first sight one may guess that the fuzzy ring configuration is only a closed type of the D1 brane,  just like the fuzzy sphere is that of the spherical D2 brane [3].   However, I will show that this is not the fact.  

  The Born-Infeld action of the $D_P$ brane we will use has the form [12]
$$  S_{D_P} = -T_p \int d^{p+1}\xi e^{-\phi} \sqrt{-\det \big(P[G]_{ab} + \lambda F_{ab} \big)},  \eqno{(4.1)}$$
where $T_p$ is the tension of the $D_P$-brane.  In the Melvin magnetic tube background (3.5) the $D1$ action becomes
$$ S_{D1} = - T_1 \int dt d\phi ~ \Lambda^{-3/4}\sqrt{-\det \Bigg(
  \begin{array}{cc}
    -\Lambda^{1/2}  & 0  \\
    0 & \Lambda^{-1/2}r^2 
  \end{array}}\Bigg) . \eqno{(4.1)}$$
The potential for this system is
$$V_{D1}= 2\pi T_1 r (1+B^2 r^2)^{-3/4}. \eqno{(4.2)}$$
The potential is zero at $r=0$ and at $r\rightarrow \infty$.  It has a local maximum at finite r.  However, the potential does not have a local minimum.   Thus we conclude that the ring configuration could not be formed from the closed D1 brane. 

To solve the problem let us consider the D2 action in the Melvin magnetic tube background (3.5)
$$ S_{D2} = - T_2 \int dt \int _0^{2\pi }d\phi \int_r^{r+\delta}dr  ~ \Lambda^{-3/4}\sqrt{-\det \Bigg(
  \begin{array}{ccc}
    -\Lambda^{1/2}  & 0&0  \\
    0 & \Lambda^{1/2}& -B r \lambda M \Lambda^{-2}\\
   0&B r \lambda M \Lambda^{-2} & \Lambda^{-1/2}r^2
  \end{array}}\Bigg)  , \eqno{(4.3)}$$
in which M is the number of magnetic flux and we have taken the integration of r from $r$ to $r +\delta$, in which $\delta$ is small.  Choosing $\delta = 2\pi\ell_s$, which is the string length scale,  the potential becomes
$$V_{ring}= 2\pi T_2~ r~ \delta ~ \sqrt{{1\over 1+B^2 r^2}\left({1+ {\lambda^2 M^2 B^2\over \Lambda^4}}\right)} = 2\pi T_1~ r~ \sqrt{{1\over 1+B^2 r^2}\left({1+ {\lambda^2 M^2 B^2\over \Lambda^4}}\right)}, \eqno{(4.4)}$$
as we have the relation  $T_2 \times 2\pi\ell_s = T_1$ in the equation (1.5).  This potential has the following properties
$$V_{ring}\rightarrow  {2\pi T_1 \over B}\left(1 - {\lambda^2 M^2 \over 2B^2r^2}\right),   ~~~~~~as~~  r \gg 1, ~~~~~\eqno{(4.5)}$$
$$V_{ring}\rightarrow  2\pi T_1 ~r~\sqrt{(1+\lambda^2 M^2 B^2)}\left(1 - \left({1\over2}B^2 + {2\lambda^2 M^2 B^4\over 1+\lambda^2 M^2 B^2}\right)~r^2\right),  ~~~~as~~  r \ll 1. ~~~~~\eqno{(4.6)}$$
The potential is zero at $r=0$ and approaches to $2\pi T_1 /B$  at $r\rightarrow \infty$.  It has a local maximum at finite $r_{*1}$ and also has a local minimum at another finite $r_{\ast 2}$, as shown in figure 2.  Therefore the stable solution may be regarded as a ring with radius $r_{* 2}$.  
\vspace{1cm}

\hfil\scalebox{1}{\includegraphics{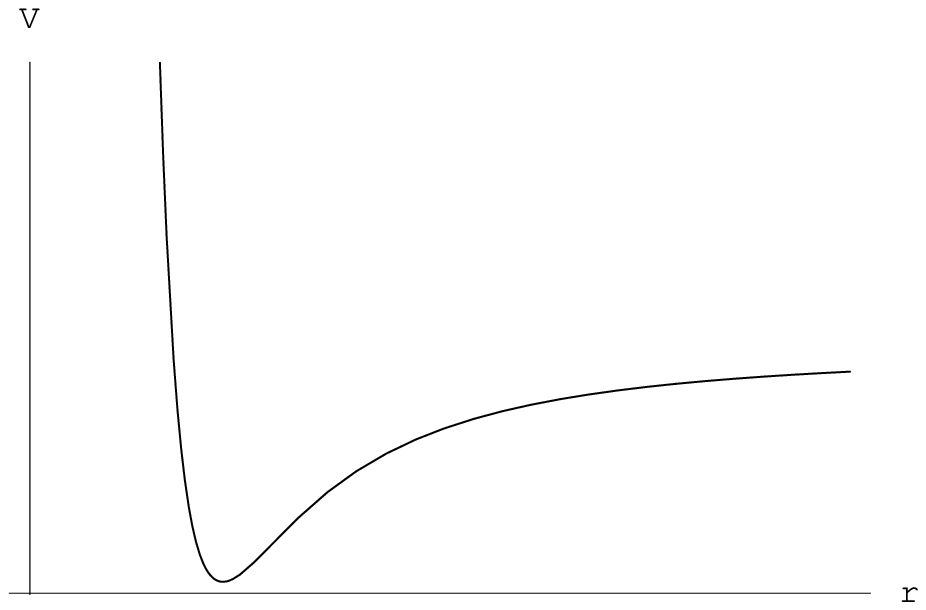}}\hfil\\
{\it ~~~Fig.2. Radius dependence of  the potential  (4.4).   This potential has a minimum at which ring configurations become stable.}
\vspace{1cm}

\section  {Discussion}

In this paper we try to find the fuzzy ring solution in the  D-branes theory.  We use the Myers T-dual nonabelin Born-Infeld action [3], which describes the M D0-branes system, to find the solutions in the background of  D6-branes and Melvin magnetic tube field.   In the  D6-Branes background we see that, besides the  fuzzy sphere solution that found by Hyakutake [11], there has also ring solutions.   These solution are formed  by the gravitational dielectric effect.    We have also seen that the fuzzy ring solution has less energy then that of the fuzzy sphere.   Thus the fuzzy sphere is metastable and would decay to the fuzzy ring configuration. In the Melvin magnetic tube field background we see that there does not exist fuzzy sphere while the fuzzy ring configuration may be formed by the magnetic  dielectric effect.  We also extend the investigation to the nonstatic system.    The results show that the $D_0$  propagating in the D6-branes and magnetic tube field background may expand into a rotating fuzzy ring.   The static solution is just the stationary state in the rotating solution.   We also use the Dirac-Born-Infeld action [12] to investigate the problems of  constructing  the ring from the D-branes.  It is seen that the fuzzy ring configuration could not be constructed from the  closed type of the D1 brane.   

  The furthermore physical properties of the fuzzy ring solution, such as its stability, the radiation during rotating, ...etc. [16], is remained to be investigated.   In this paper we have considered only the Boson Hamiltonian.   The Fermion part can be obtained by the supersymmetrization.    The interaction between a pair of the fuzzy ring and other dynamics of the fuzzy ring can be investigated from the complete Hamiltonian.  These problems remain to be studied.

\newpage
\begin{center} {\large \bf  References} \end{center}
\begin{enumerate}
\item J.~Polchinski, Phys. Rev. Lett.  75 (1995) 4724 [hep-th/9510017].
\item  W.~Taylor, ``Lectures on D-branes, gauge theory and M(atrices),'' at 2nd Trieste Conference on Duality in String Theory, Trieste, Italy, 16-20 Jun 1997 [hep-th/9801182].
\item R.C. Myers, ``Dielectric-Branes'',  JHEP 9912 (1999) 022 [hep-th/0205185].  
\item D. Mateos and P. K. Townsend, ``Supertubes'', Phys. Rev. Lett. 87 (2001) 011602 [hep-th/0103030];\\
D. Bak, Ki-Myeong Lee, ``Noncommutative Supersymmetric Tubes'',  Phys. Lett. B509 (2001) 168 [hep-th/0103148];\\
Y. Hyakutake, ``Torus-like Dielectric D2-brane'' ,  JHEP 0105 (2001) 013 [ hep-th/0103146];  ``Notes on the Construction of the D2-brane from Multiple D0-branes'',  Nucl. Phys.  B675 (2003) 241 [hep-th/0302190];\\
Y. Hyakutake and N. Ohta, ``Supertubes and Supercurves from M-Ribbons,''
Phys. Lett. B539  (2002) 153 [hep-th/0204161].

\item M. A. Melvin, ``Pure magnetic and electric geons,'' Phys. Lett. 8 (1964) 65.
\item G.~W.~Gibbons and K.~Maeda, ``Black holes and membranes in higher dimensional theories with dilaton fields,'' Nucl.\ Phys.\ B298 (1988) 741.
\item  J.~G.~Russo and A.~A.~Tseytlin, ``Exactly solvable string models of curved space-time backgrounds,'' Nucl.\ Phys.\ B449 (1995) 91 [hep-th/9502038]; ``Magnetic flux tube models in superstring theory,'' Nucl.\ Phys.\ B461 (1996) 131 [hep-th/9508068].
\item F.~Dowker, J.~P.~Gauntlett, D.~A.~Kastor and J.~Traschen, ``The decay of magnetic fields in Kaluza-Klein theory,'' Phys.\ Rev.\ D52 (1995) 6929 [hep-th/9507143];\\
M.~S.~Costa and M.~Gutperle, ``The Kaluza-Klein Melvin solution in M-theory,'' JHEP 0103 (2001) 027 [hep-th/0012072].
\item  A. Adams, J. Polchinski and E. Silverstein,  ``Don't Panic! Closed String Tachyons in ALE Spacetimes,''  JHEP 0110 (2001) 029 [hep-th/0108075]; \\
J.~R.~David, M.~Gutperle, M.~Headrick and S.~Minwalla, ``Closed string tachyon condensation on twisted circles,''  JHEP 0202 (2002) 041 [hep-th/0111212].
\item M.~Gutperle and A.~Strominger, ``Fluxbranes in string theory,''
JHEP 0106 (2001) 035 [hep-th/0104136]; \\
R.~Emparan and M.~Gutperle, ``From p-branes to fluxbranes and back,'' JHEP 0112 (2001) 023 [hep-th/0111177];\\
M.~S.~Costa, C.~A.~Herdeiro and L.~Cornalba, ``Flux-branes and the dielectric effect in string theory,'' Nucl.\ Phys.\ B619 (2001) 155. [hep-th/0105023];\\
 T.~Takayanagi and T.~Uesugi,  ``D-branes in Melvin background,'' JHEP 0111 (2001) 036 [hep-th/0110200].
\item Y. Hyakutake, ``Gravitational Dielectric Effect and Myers Effect,''  [hep-th/0403026].
\item  R. G. Leigh, ``Dirac-Born-Infeld Action From Dirichlet Sigma Model'', 
Mod. Phys. Lett.  A4 (1989) 2767.
\item Wung-Hong Huang,``Fluxbrane in Melvin Background: Membrane Matrix Approach,'' [hep-th/0403181].
\item B. de Wit, K. Peeters, J. Plefka, ``Superspace Geometry for Supermembrane  Backgrounds,''  Nucl.Phys. B532 (1998) 99, [hep-th/9803209].
\item  M. Alishahiha, Y. Oz, M. M. Sheikh-Jabbari, ``Supergravity and Large N Noncommutative Field Theories,'' JHEP 9911 (1999) 007  [hep-th/9909215];  M. Alishahiha, H. Ita, and Y. Oz,``Graviton Scattering on D6 Branes with B Fields,'' JHEP 0006 (2000) 002  [hep-th/0004011].
\item T. Harmark and K. G. Savvidy ,``Ramond-Ramond- Field Radiation from Rotating Ellipsoidal Membranes,''  Nucl.Phys. B585 (2000) 567, [hep-th/0002157].

\end{enumerate}
\end{document}